\documentclass[fleqn,usenatbib]{mnras}

\usepackage{newtxtext,newtxmath}

\usepackage{siunitx}
\usepackage[T1]{fontenc}

\DeclareRobustCommand{\VAN}[3]{#2}
\let\VANthebibliography\thebibliography
\def\thebibliography{\DeclareRobustCommand{\VAN}[3]{##3}\VANthebibliography}

\usepackage{graphicx}	
\usepackage{amsmath}	
\usepackage{subcaption}
\usepackage{mwe}

\title{Evidence for Rapid Variability at High Energies in GRBs}

\author[Aldrich \& Nemiroff]
{E. Casey Aldrich,$^{1}$\thanks{E-mail: ecaldric@mtu.edu} 
and 
{Robert J. Nemiroff}$^{1}$\thanks{E-mail: nemiroff@mtu.edu}
\\
$^{1}$Dept. of Physics, Michigan Technological University, Houghton, MI, USA\\
}

\date{Accepted XXX. Received YYY; in original form ZZZ}

\pubyear{2024}

\begin{document}
\label{firstpage}
\pagerange{\pageref{firstpage}--\pageref{lastpage}}
\maketitle

\begin{abstract}
Intrinsic variability was searched for in arrival times of six gamma-ray bursts (GRBs) at high energies -- between 30 MeV and 2 GeV -- detected by the Fermi satellite's Large Area Telescope (LAT). The GRBs were selected from the Fermi LAT catalog with preference for events with numerous photons, a strong initial pulse, and measured redshifts. Three long GRBs and three short GRBs were selected and tested. Two different variability-detection algorithms were deployed, one counting photons in pairs, and the other multiplying time gaps between photons. In both tests, a real GRB was compared to 1000 Monte-Carlo versions of itself smoothed over a wide range of different timescales. The minimum detected variability timescales for long bursts (GRB 080916C, GRB 090926A, GRB 131108A) was found to be (0.005, 10.0, 10.0) seconds for the photon pair test and (2.0, 20.0, 10.0) seconds for the time-gap multiplication test. Additionally, the minimum detected variability timescales for the short bursts (GRB 090510, GRB 140619B, GRB 160709A) was found to be (0.05, 0.01, 20.0) seconds for the photon pair test and (0.05, 0.01, 20.0) seconds for the gap multiplication test. Statistical uncertainties in these times are about a factor of 2. The durations of these variability timescales may be used to constrain the geometry, dynamics, speed, cosmological dispersion, Lorentz-invariance violations, weak equivalence principle violations, and GRB models. 
\end{abstract}




\section{Introduction}

Gamma-ray bursts (GRBs) are among the most distant known transient events, with redshifts typically greater than one \citep{2018Ap&SS.363..223Z} and as great as (estimated) redshift 9 \citep{2011ApJ...736....7C}. A comprehensive living list of GRBs with both measured and estimated redshifts is provided by \citet{GreinerWebCat}. The significant distance to GRBs can provide valuable scientific insights into various aspects of the young universe, such as early stellar evolution, star formation rates, and the physics of the early cosmological epochs. For a good review of GRBs, see \citeauthor{2018grb..book.....L} (\citeyear{2018grb..book.....L}).

Along with Fast Radio Bursts, GRBs are among the shortest-duration astronomical events known. GRBs have measured durations ranging from milliseconds to hours at gamma-ray energies  \citep{1995ApJ...439..542N, 2012MNRAS.425L..32M}. GRBs are often characterized by constituent pulses that are dominated by a relatively rapid rise and slow decay, which have been extensively studied \citep{1981Ap&SS..75...15D, 1996ApJ...459..393N, 2002ApJ...566..210R, 2012MNRAS.419.1650N}. Broad temporal characteristics of GRBs are important for providing insights into the internal physics and kinematics of these phenomena \citep{2004RvMP...76.1143P}.

GRBs are commonly categorized as either long or short, based on their duration and spectral properties \citep{1993ApJ...413L.101K, 2015MNRAS.451..126S}. Long GRBs typically last for more than 2 seconds and exhibit relatively soft spectra. A prevailing progenitor model for these events involves a core-collapse supernova that occurs when a massive star depletes the fusible elements in its core \citep{2006ARA&A..44..507W}. In contrast, short GRBs generally have durations of less than 2 seconds and are characterized by harder spectra. The dominant progenitor model for short GRBs posits that they result from a kilonova, which is associated with the merger of two neutron stars that eventually form a black hole \citep{1989Natur.340..126E}.

The contemporary GRB detector most sensitive to individual photons near 100 MeV (hereafter called ``high-energy") is the Large Area Telescope (LAT, \citealt{2009ApJ...697.1071A}) aboard the Fermi Gamma-Ray Space Telescope (Fermi). The Fermi LAT primarily detects gamma-ray photons between 20 MeV and 10 GeV. A summary of GRB events can be found in the Fermi LAT's First and Second GRB Catalogs \citep{2013ApJS..209...11A, 2019ApJ...878...52A}. In the decade from mid-2008 to mid-2018, the Fermi LAT recorded 186 GRBs with a significant excess of photons \citep{2019ApJ...878...52A}. In comparison, during this time, the Fermi Gamma-ray Burst Monitor (GBM), which detects gamma-rays between 8 keV and 40 MeV (hereafter called ``low-energy") -- but mostly at the lower end -- triggered on 2357 GRBs \citep{2019ApJ...878...52A}, indicating that relatively few GRBs emit a significant excess of LAT-detectable high-energy photons. However, in contrast with the GBM, individual photons detected by the LAT that are time and direction related to a GRB have a high confidence of being actually GRB photons -- and not from a pervasive background. 

Rapid variability in the prompt gamma-ray emission of GRBs has been searched for many times and with different methods, but mostly at the lower photon energies detectable by GBM. Millisecond variability was detected at near the 100 keV energy scale \citep{1992Natur.359..217B, 1999PhRvL..82.4964S}. \citet{2000ApJ...537..264W} searched the initial one second of many BATSE GRBs in the energy range 25 keV to 1 MeV and found that 30 percent showed a rise-time variability of less than 1 ms. None of these GRBs, though, had a known redshift. 

\citet{2012MNRAS.425L..32M} analyzed pulses that occurred in both short GRBs and long GRBs detected by the GBM with a wavelet analysis and found variability timescales between 0.001 and 0.01 seconds as well as a temporal offset between short and long GRBs. Subsequently, \citet{2013MNRAS.432..857M} studied GBM bursts using a wavelet analysis and found several short GRBs that have a variability timescale between 0.001 and 0.01 seconds, as well as several long GRBs with variability timescales from 0.01 seconds to 1 second. Variability timescale has also been used to estimate the size of the emitting region of GRBs \citep{2014Sci...343...42A}. Previous results have also pointed to there being a correlation between variability timescales and burst duration, with the timescales of long GRBs being different from those of short GRBs \citep{2013MNRAS.432..857M}. 

The variability of GRBs in the LAT's high-energy gamma-ray range is less well studied. Even so, the timescale of GRB variability at LAT energies has been used by \citet{2012MNRAS.421..525H} to constrain engine factors like $\gamma-\gamma$ opacity and the Lorentz factor of outflows. The same study also found that it is not unusual for these timescales to be on the order of a few milliseconds. \citet{2017A&A...606A..93Y} also calculates the variability timescale of a long GRB, GRB 090926A, to be about 10 milliseconds at high energies before using this value to estimate the Lorentz factor of its beam. 

A common method for estimating variability at LAT energies is to measure the widths of GRB constituent pulses \citep{2012MNRAS.421..525H}. In general, GRB pulses are narrower in time at higher energies than lower energies \citep{2000ApJ...544..805N}, which empirically indicates that GRB variability timescales may also appear shorter at higher gamma-ray energies. GRB timescales on the order of one millisecond have been detected in LAT data at the GeV energy scale using an auto-correlation method in photon arrival times \citep{2012PhRvL.108w1103N}. 

A favored mechanism for gamma-ray production in GRBs is synchrotron radiation created by high-energy electrons being accelerated by forward shocks with turbulent magnetic fields \citep{1998ApJ...506L..23P}. However, at the high energies detected by the LAT, photons from inverse Compton scattering -- where low energy photons are upscattered by relativistic electrons -- may become important. The underlying physics of gamma-ray photon creation in GRBs remains a topic of study, and it is not clear the LAT-detected photons are produced, in general, by the same physics that produce the lower-energy GBM-detected photons \citep{2011A&A...526A.110D}. It is therefore possible that different photon creation mechanisms, if active, create different timescales for photon bunching in high and low energy regimes.

The combination of great distance and rapid variability makes GRBs a valuable tool for exploring the potentially temporally-dispersive properties of the universe between Earth and the GRB. Theoretically, processes that could cause temporally disperse GRB photons include Lorentz-invariance violations \citep{2002MPLA...17.1025S}, weak equivalence principle violations \citep{2015ApJ...810..121G, 2016ApJ...821L...2N, 2019arXiv190305688T}, and potentially physical mechanisms associated with cosmologically-distributed dark matter, dark energy, or vacuum fluctuations \citep{2012PhRvL.108w1103N}.

Following an unexpected and informal indication of sub-pulse bunching in the arrival times of high-energy photons in GRB 080916C, a formal search was initiated to explore the reality of this bunching and search for potentially similar bunchings at high energy in other LAT-detected GRBs. The five other GRBs studied were chosen for a relatively high number of LAT-detected photons, a strong initial pulse, and recorded spectroscopic redshifts -- the latter to strengthen potential cosmological implications. This study settled on three long GRBs (GRB 080916C, GRB 090926A, and GRB 131108A) and three short GRBs (GRB 090510A, GRB 140619B, and GRB 160709A), with all except GRB 140619B and GRB 160709A having spectroscopic redshifts. 

Section 2 of this article provides a description of the methodology employed to explore short duration bunching, while Section 3 details the results obtained through analysis. In Section 4, the results are summarized and discussed in the context of the broader research objectives.

\section{Methods}
\label{sec:methods} 

What is the shortest timescale that really exists in a GRB? For GRBs that are obviously sectioned into constituent pulses, are there inherent timescales shorter than that of the pulses? Mathematically, the shortest timescale measurable is the time between consecutive photon arrival times, but that is not a characteristic of the GRB because it depends on the size and sensitivity of the detector. 

This work attempts to find sub-pulse timescales inherent to GRBs by a relatively simple method -- smoothing a given GRB on a series of timescales and then looking for statistically significant bunching of photon arrival times in the original GRB as compared to its smoothed versions.

\subsection{Data Preparation}

The Fermi LAT identifies the arrival times and directions for individual high-energy photons \citep{WebSiteFermi}. For most GRBs, the LAT records no GRB-associated events. Conversely, in rare cases, the LAT records over 100 high-energy GRB-coincident events. This work, of course, focuses on the latter. The analysis presented here analyzed Fermi GRB data Pass 8 \citep{2013arXiv1303.3514A}.

Fermi LAT data generally comprises two distinct relevant entry types: "Events" and "Photons," with  capitalization here indicating data types explicitly labeled in Pass 8 files. The arrival times of both Events and Photons are measured by the Fermi LAT with a temporal accuracy of approximately 10 microseconds \citep{2009ApJ...697.1071A}. Events may include charged particles, such as electrons and protons, and photons. In contrast, entries labeled as Photons are determined from attributes and trajectories within the LAT, during the Pass 8 analysis, to be most likely photons \citep{2013arXiv1303.3514A}.

In the current analysis, both Event and Photon entries falling within a 3-degree angular error circle, and nearly time-coincident with the GRB trigger time, are considered to be genuine GRB photons. Drawing from past experience with LAT-detected GRBs, 3 degrees was chosen to be large enough to capture most GRB LAT-detected photons, but small enough to exclude most of the irrelevant background. While a generic Event occurring at a random location and time may have an ambiguous origin, Events that are both location and time coincident with a known GRB have a much higher likelihood of being GRB-related. This principle is examined in detail in \citet{2022RNAAS...6..237A}.

The total number of Photons and Events for the different bursts used in this analysis were as follows, GRB 080916C: 321, GRB 090926A: 404, GRB 131108A: 252, GRB 090510: 287, GRB 140619B: 31, and GRB 160709A: 76.

\subsection{Smoothing Light Curves}

To determine the shortest timescale $t_{min}$ with statistical significance in a given GRB or a portion of a GRB in LAT data, each analyzed GRB was smoothed over a range of timescales $t_{smooth}$. The smoothing method implemented here involved assuming a kernel density function (KDF) \citep{KDFarticle}, which assigns weights to photons arriving at various positions within a sliding time-based kernel window. This sliding window is advanced along the GRB, being centered on one photon at a time with a temporal width of $t_{smooth}$. Summing these KDFs up creates a kernel density estimation (KDE) \citep{KDEbook}, which is utilized to generate Monte-Carlo (MC) versions of the GRB smoothed over a specific timescale. This procedure creates an effective brightness curve of the given GRB. 

The primary KDF smoothing window utilized was Gaussian in shape. Therefore, each photon's arrival time was surrounded by a Gaussian KDF with a temporal width $t_{smooth}$. These KDFs were then summed to create the KDE \citep{CodeCiteTriveri}. The KDEs of the GRBs were constructed using all LAT recorded photons with an arrival time within the first 1000 seconds following the trigger time. A Monte-Carlo (MC) simulation was used to set up the statistical tests for these GRBs, with a number of random photons thrown under the selected section of the brightness curve equivalent to the number of photons in the original burst's time range. MC simulations of 1000 realizations were performed to search for photon bunching between the actual arrival times of photons in the GRB and the arrival times under the Gaussian-smoothed GRB. 

To investigate potential bunching on multiple timescales, both $t_{smooth}$ and a potential bunching scale $\Delta t$ were assigned a range of values, logarithmically from $10^{-6}$ sec to 50 seconds, increasing each time by a roughly a factor of 2. Specifically, the discrete timescales employed were, in seconds, $10^{-6}$, 2 x $10^{-6}$, 5 x $10^{-6}$, $10^{-5}$, 2 x $10^{-5}$, 5 x $10^{-5}$, $10^{-4}$, 2 x $10^{-4}$, 5 x $10^{-4}$, $10^{-3}$, 2 x $10^{-3}$, 5 x $10^{-3}$, 0.01, 0.02, 0.05, 0.10, 0.20, 0.50, 1, 2, 5, 10, 20, and 50.

\subsection{Counting Photons in Pairs}

The first method utilized here for identifying the bunching of GRB photons involved the cumulative counting of photons that were part of a pair of photons arriving within a time gap $\Delta t$ or less. This method, referred to as a ``pair analysis," entailed evaluating each GRB photon sequentially. Whenever a photon arrived within a predetermined time difference $\Delta t$ of another photon, the tally $R$ for photon pairs was incremented. 

In total, 1000 MC simulations were created for each GRB. For each $\Delta t$, the fraction of the number of MC runs where the number of bunched photons in the MC smoothed GRBs was greater than or equal to that of the real GRB was computed. The lowest fraction found from all $\Delta t$ values was noted. This proceeded for all $t_{smooth}$ values.

A significant concentration of bunched photons was deemed present if the fraction of MC GRBs having more than or the same number of bunched pairs as the real GRB was sufficiently small -- occurring in less than 1 in 370 of the MC smoothed GRBs for some $\Delta t$. This threshold is derived from an analogous $3 \sigma$ threshold in a normal distribution. The second smallest $t_{smooth}$ value that shows this rare a bunching for any $\Delta t$ was declared to be $t_{min}$ for that GRB, the estimated timescale of minimum detectable variability. 

For clarity, an example is given here. Suppose the arrival times of 100 photons for GRB 000001A were recorded by the Fermi LAT. An analysis showed that $R = 10$ photons were found to be part of a pair separated by $\Delta t < 0.1$ second, while 50 photons by $\Delta t < 10$ sec. Now, with $t_{smooth} = $ 1.0 sec, this GRB's photons are reshuffled in a way that preserves the GRB light-curve shape smoothed over 1.0 second. It is now found for this MC GRB that 4 photons are separated by $\Delta t < 0.1$, and 52 by $\Delta t < 10$ sec. Then a total of 1000 MC smoothed GRBs were digitally created, also with $t_{smooth} = $ 1.0 sec. Considering all 1000 MC GRBs, only 2 were found to have more bunching on the $\Delta t < 0.1$ second timescale than the real GRB, while 998 of them had less bunching. Additionally, on the $\Delta t =$ 10 second timescale, 400 MC GRBs had more bunching than the real GRB, and 600 had less bunching. From this it can be seen that this GRB showed significant bunching when $t_{smooth} = $ 1.0 second and $\Delta t =$ 0.1 seconds, but insignificant bunching on the $\Delta t = 10$ second timescale. It could then be concluded that GRB 000001A has significant bunching not on the $\Delta t$ timescale of 0.1 second, but at the $t_{smooth}$ timescale of 1.0 sec. 

Note that the $2 \sigma$ and $3 \sigma$ thresholds do not take into account the number of $\Delta t$ trials. Each $\Delta t$ value analyzed can be considered a new trial for each $t_{smooth}$. However, numerous $t_{smooth}$ values satisfied the $\sigma$ thresholds, and consecutive $\Delta t$ values are not fully statistically independent. Therefore, the $\sigma$ values considered here are more suggestive of statistical significance than actual formal delimiters. To be conservative, the {\it second} lowest $t_{smooth}$ value that satisfies the $3 \sigma$ threshold was considered more indicative of $t_{min}$, with an error bar spanning the duration between the first and second lowest $t_{smooth}$ with pair fraction below $3 \sigma$.

\subsection{Multiplying Time Gaps}

A second method used here to detect photon bunching in GRBs is multiplying time gaps. After choosing a GRB and determining practical start and end times for arriving events, the gaps in times between photon arrival times were calculated. Now adding all of these time gaps together will just give ($t_{end}$ - $t_{start}$) no matter how bunched these photons are, but multiplying the gaps together gives a result $S$ that depends on how bunched the photons are. The largest number possible from multiplying all of the time gaps together, $S_{max}$, occurs when the photons are uniformly spread across the detection time interval. Any deviation from this uniformity will result in a lower $S$ value. In general, the more bunched the photon arrival times, the smaller $S$ becomes relative to $S_{max}$. The variable $S$ has analogies to both physical and informational entropy. 

The analysis procedure of time-gap multiplication was as follows. First, for a given real GRB, the value of $S$ was computed. Next, the sets of 1000 MC smoothed GRBs were recalled from the pair bunching analysis, each set smoothed over a timescale $t_{smooth}$. To be clear, these recalled MC-smoothed GRBs each had an assumed KDF -- assumed Gaussian -- from which smoothed KDEs (light curves) were generated.

For each MC GRB smoothed by $t_{smooth}$, a value of $S_{smooth}$ was computed by multiplying together the gaps in times between all arrival times. To avoid numerical overflows, the cumulative product was computed in a logarithmic space -- by taking the natural logarithm of all of the time gaps and adding them together. If the value of $S$ for the real GRB was less than 1 in 370 of the $S_{smooth}$ values for the MC smoothed GRBs, then that GRB was deemed to have significant photon bunching over this $t_{smooth}$. This procedure was repeated for each $t_{smooth}$ value. Again, the second lowest value of $t_{smooth}$ where $S$ occurred less frequently than 1 in 370 $S_{smooth}$ values was deemed $t_{min}$. The error in $t_{min}$ was considered to be the difference in time between the two lowest $t_{smooth}$ values where $S$ occurred less frequently than 1 in 370 $S_{smooth}$ values.

\section{Results}

The methods and analyses of the previous section were applied to the six GRBs selected. The Fermi LAT photons used in this work for the GRBs analyzed are shown in Figures 1 and 2.

Before presenting the results of the main bunching analyses, two preliminary results are presented that bolster known attributes of GRBs. The first is that at high energies, the onset of the GRB was really quite sudden and not just the flux peak of a longer time-symmetric train of photons. This is apparent for all six GRBs studied by inspection of Figures 1 and 2, including both long and short GRBs. Although Figure 1 shows that there are photons that arrive well before trigger, they are noticeably less numerous than the number of photons that arrive quite late. Although this may indicate that the very early pre-trigger photons are from an unrelated cosmic background, it more surely indicates that GRBs have high-energy photons that keep arriving many seconds after trigger, even for ``short" GRBs.

A second preliminary result is also visible in Figures 1 and 2. There, near the time of GRBs, the LAT pass-8 data listed as "Events", shown as red dots, arrive with similar times to the LAT pass-8 data listed as "Photons", shown with blue dots. This was described previously by \citet{2022RNAAS...6..237A}. To be clear, Figures 1 and 2 show that LAT Events near in time and angular location to Photons are overwhelmingly likely, per Event, to signify real GRB photons.

Additionally, inspection of the photon arrival times in Figure 2 indicates that the initial pulse for the long GRBs, at high energies, has a duration of about 20 seconds, while the initial pulse for the short GRBs, over the same high energies, has a duration of about 2 seconds. 

The main results of this analysis are shown in Figures 3 and 4, with Figure 3 summarizing results from the cumulative pair analysis, and Figure 4 summarizing results from the multiplying time gaps analysis. The $x$-axes of both plots label the smoothing time $t_{smooth}$ used to digitally generate the smoothed MC bursts. The $y$-axes label the fraction of smoothed GRBs that exhibited the same or greater number of bunched photons than the real burst at any $\Delta t$. A very low fraction at a given $t_{smooth}$ indicates that the bunching in the real burst is significant at that smoothing timescale.

Several things become clear from inspection of Figures 3 and 4. First, typically, when the smoothing timescale $t_{smooth}$ is small, a MC smoothed burst is effectively unsmoothed and so has a similar number of photons in pairs as the real GRB. This is why the fraction is near unity on the left of the plots in Figures 3 and 4.

Conversely, when the smoothing timescale $t_{smooth}$ is large -- say near the duration of the GRB, then MC versions of the GRB are smoothed by having their photons randomized over much of the GRB. Therefore, almost any structure in the GRB is smoothed away, and the number of close MC pairs drops to below that of the real GRB. This is why the fraction is near zero on the right of the plots in Figures 3 and 4.

The dashed pink and black lines in Figures 3 and 4 indicate where the results become significant at the $2 \sigma$ and $3 \sigma$ levels respectively. The second smallest $t_{smooth}$ value with $3 \sigma$ significance is then referred to here as the GRB's minimum detectable timescale from this data analysis: $t_{min}$. For the long GRBs, GRB 080916C, GRB 090926A, and GRB 131108A, $t_{min}$ was found to be 0.005, 10, and 10 seconds  respectively. For the short GRBs studied, GRB 090510, GRB 140619B, and GRB 160709A, $t_{min}$ was found to be 0.05, 0.01, and 20 seconds respectively. Since 20 seconds is on the order of this short GRB's LAT duration \citep{2019ApJ...878...52A}, this $t_{min}$ just means that no reportable inter-GRB photon bunching was detected.

The pair analysis results were checked by utilizing a uniformly flat KDF in place of the Gaussian. Timescales for all the GRBs were usually slightly longer with this kernel. For the long GRBs, $t_{min} = $ 0.01 seconds for GRB 080916C, $t_{min} = $ 20 seconds for GRB 090926A, and $t_{min} = $ 10 seconds for GRB 131108A. For the short GRBs, $t_{min} = $ 0.2 seconds for GRB 090510, $t_{min} = $ 0.05 seconds for GRB 140619B, and $t_{min} = $ 20 seconds for GRB 160709A. Since each kernel has its own attributes, the differences in $t_{min}$ values should be attributed to inherent numerical variance in the KDE methodology.

The plots in Figure 4 show the results of the time-gap multiplication analysis. In sum, for the long GRBs, GRB 080916C, GRB 090926A, and GRB 131108A, $t_{min}$ was found to be 2, 20, and 10 seconds respectively. For the short GRBs studied, GRB 090510, GRB 140619B, and GRB 160709A, $t_{min}$ was found to be 0.05, 0.01, and 20 seconds respectively.

These time-gap multiplication analysis results were also checked by utilizing a uniformly flat KDF in place of the Gaussian. Similarly, timescales for the long GRBs were slightly shifted with this kernel, with $t_{min} = $ 0.5 seconds for GRB 080916C, $t_{min} = $ 20 seconds for GRB 090926A, and $t_{min} = $ 10 seconds for GRB 131108A. The timescales for the short GRBs studied where also shifted, with $t_{min} = $ 0.2 seconds for GRB 090510, $t_{min} = $ 0.05 seconds for GRB 140619B, and $t_{min} = $ 50 seconds for GRB 160709A. Again, differences with the Gaussian kernel can be attributed to inherent numerical variance.

The terms $2 \sigma$ and $3 \sigma$ are used here to indicate levels of significance in familiar terms, but are not meant to indicate that the distributions of $R$ or $S$ are Gaussian. Classically, the numerical $\sigma$ threshold, for example 1 in 370 for the $3 \sigma$ designation, implies significance could be obtained on either side of a two-sided Gaussian distribution curve. However in this work, significance was only searched for on one side of the distribution: the unusual bunching side and not the unusual smoothness side of photon arrival times. 

\begin{figure*}
    \centering
    \begin{subfigure}[b]{0.475\textwidth}
        \centering
        \includegraphics[width=\textwidth]{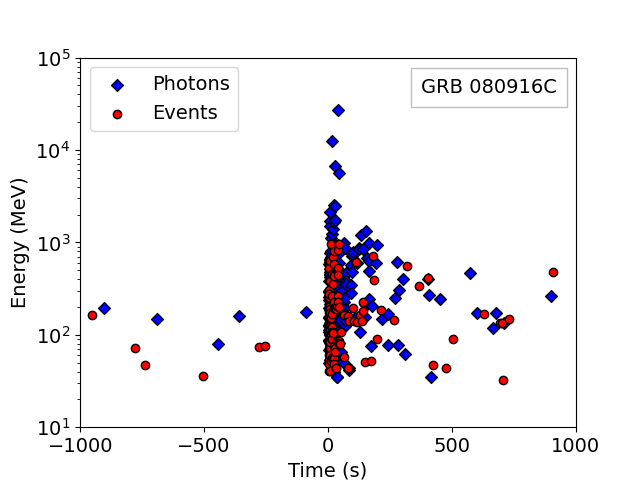}
        \caption[]%
        {{\small GRB 080916C}}    
        \label{080916Cphotons}
    \end{subfigure}
    \hfill
    \begin{subfigure}[b]{0.475\textwidth}  
        \centering 
        \includegraphics[width=\textwidth]{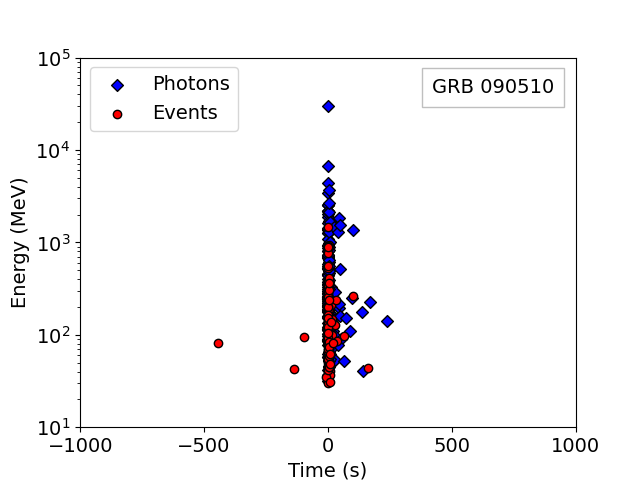}
        \caption[]%
        {{\small GRB 090510}}    
        \label{090510photons}
    \end{subfigure}
    
    \begin{subfigure}[b]{0.475\textwidth}   
        \centering 
        \includegraphics[width=\textwidth]{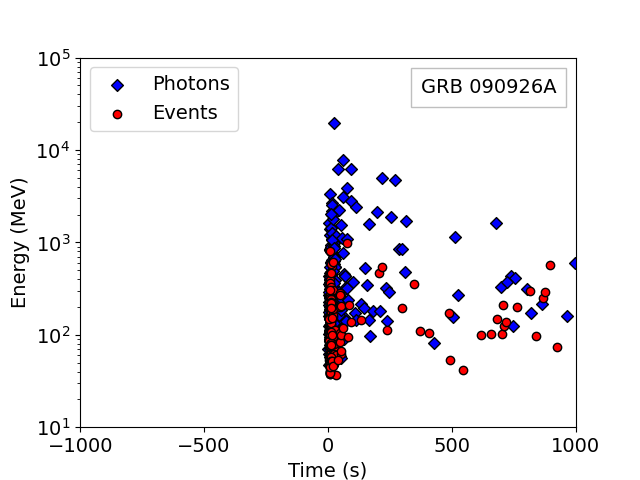}
        \caption[]%
        {{\small GRB 090926A}}    
        \label{090926Aphotons}
    \end{subfigure}
    \hfill
    \begin{subfigure}[b]{0.475\textwidth}   
        \centering 
        \includegraphics[width=\textwidth]{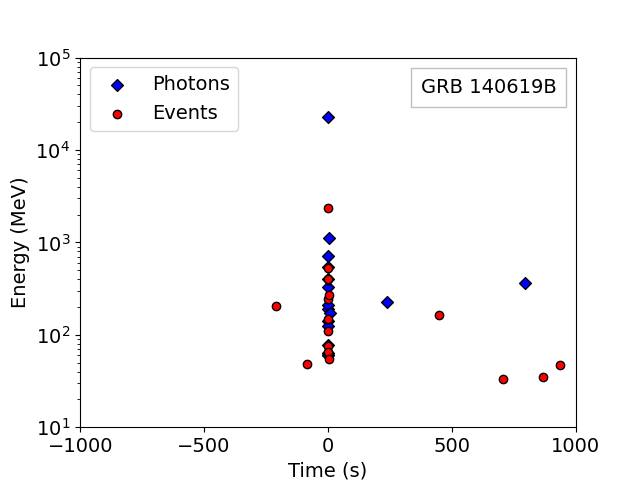}
        \caption[]%
        {{\small GRB 140619B}}    
        \label{140619Bphotons}
    \end{subfigure}
    
    \begin{subfigure}[b]{0.475\textwidth}  
        \centering 
        \includegraphics[width=\textwidth]{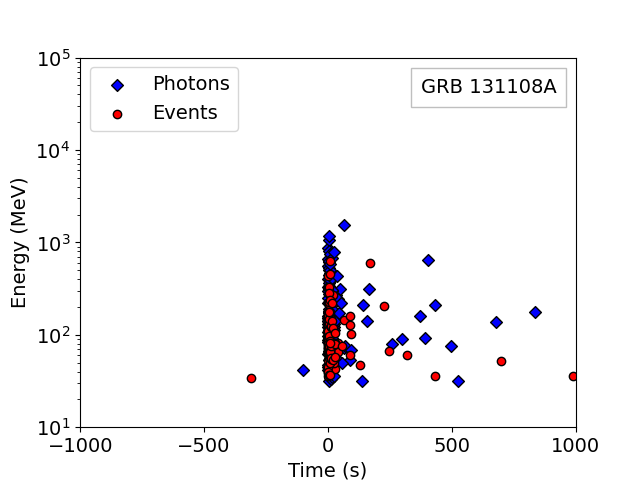}
        \caption[]%
        {{\small GRB 131108A}}    
        \label{131108Aphotons}
    \end{subfigure}
    \hfill
    \begin{subfigure}[b]{0.475\textwidth}   
        \centering 
        \includegraphics[width=\textwidth]{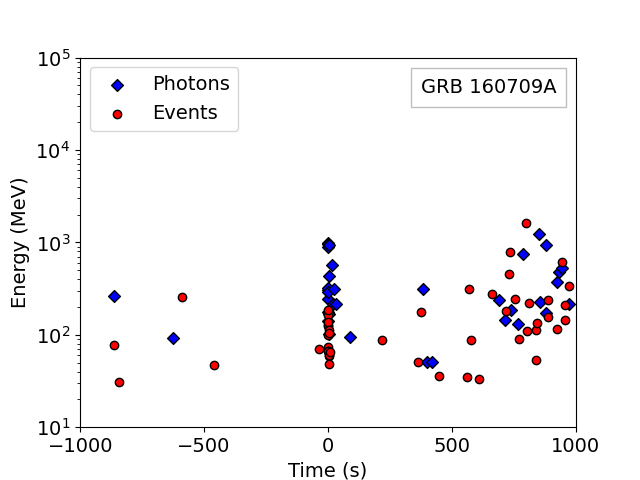}
        \caption[]%
        {{\small GRB 160709A}}    
        \label{160709Aphotons}
    \end{subfigure}
    \caption[]
    {\small Plots of photon arrival times versus photon energy for each of the six GRBs analyzed. Blue dot represent LAT events designated "Photons", while red dots represent "Events".  The left three plots are for long GRBs, while the right three are for short GRBs.  } 
    \label{GRBphotons}
\end{figure*}

\begin{figure*}
    \centering
    \begin{subfigure}[b]{0.475\textwidth}
        \centering
        \includegraphics[width=\textwidth]{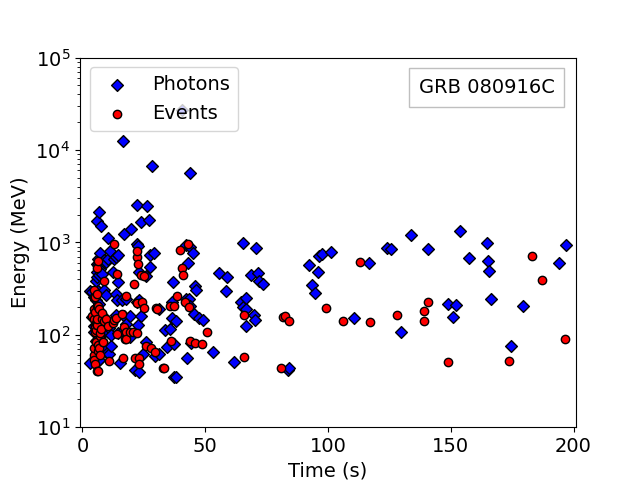}
        \caption[]%
        {{\small GRB 080916C}}    
        \label{080916Cphotonspulse}
    \end{subfigure}
    \hfill
    \begin{subfigure}[b]{0.475\textwidth}  
        \centering 
        \includegraphics[width=\textwidth]{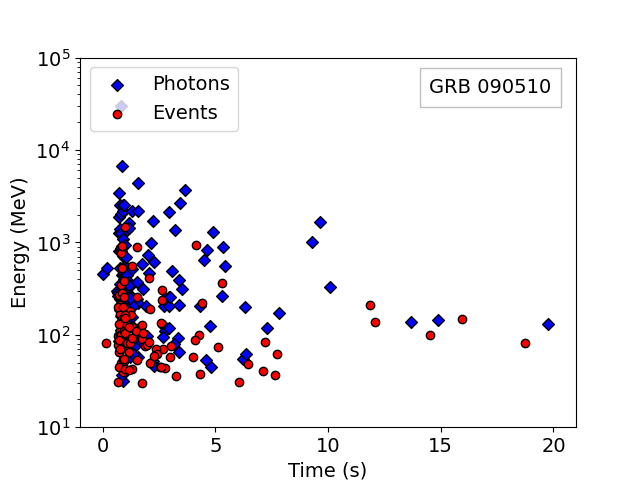}
        \caption[]%
        {{\small GRB 090510}}    
        \label{090510photonspulse}
    \end{subfigure}
    
    \begin{subfigure}[b]{0.475\textwidth}   
        \centering 
        \includegraphics[width=\textwidth]{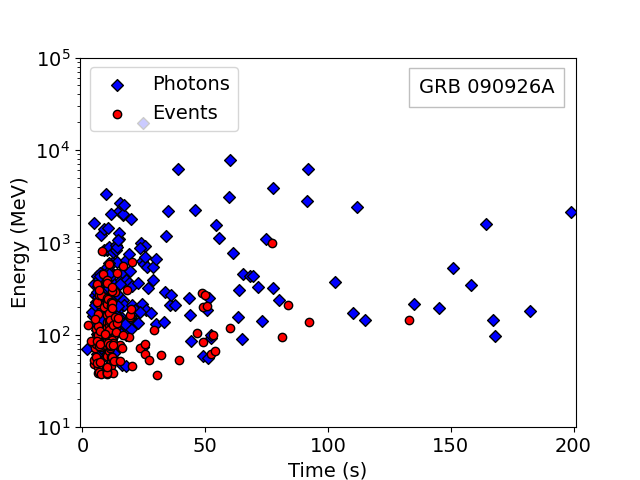}
        \caption[]%
        {{\small GRB 090926A}}    
        \label{090926AAphotonspulse}
    \end{subfigure}
    \hfill
    \begin{subfigure}[b]{0.475\textwidth}   
        \centering 
        \includegraphics[width=\textwidth]{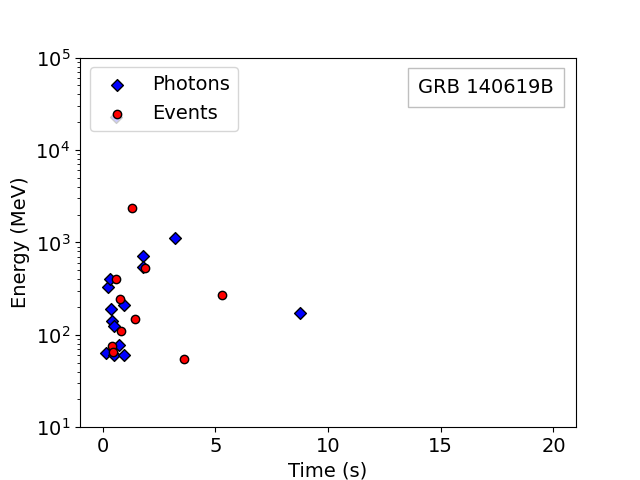}
        \caption[]%
        {{\small GRB 140619B}}    
        \label{140619Bphotonspulse}
    \end{subfigure}
    
    \begin{subfigure}[b]{0.475\textwidth}  
        \centering 
        \includegraphics[width=\textwidth]{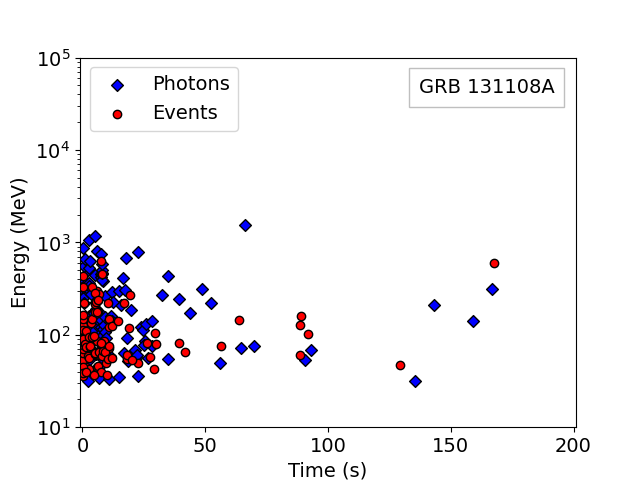}
        \caption[]%
        {{\small GRB 131108A}}    
        \label{131108Aphotonspulse}
    \end{subfigure}
    \hfill
    \begin{subfigure}[b]{0.475\textwidth}   
        \centering 
        \includegraphics[width=\textwidth]{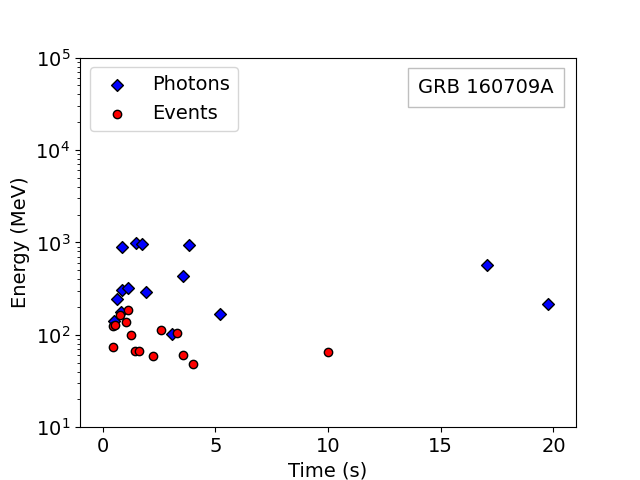}
        \caption[]%
        {{\small GRB 160709A}}    
        \label{160709Aphotonspulse}
    \end{subfigure}
    \caption[]
    {\small Plots of photon arrival times versus photon energy for each of the six GRBs analyzed as shown in Figure 1, but over a shorter temporal period near trigger to show earlier times more clearly. 
    } 
    \label{GRBphotons2}
\end{figure*}

\begin{figure*}
    \centering
    \begin{subfigure}[b]{0.475\textwidth}
        \centering
        \includegraphics[width=\textwidth]{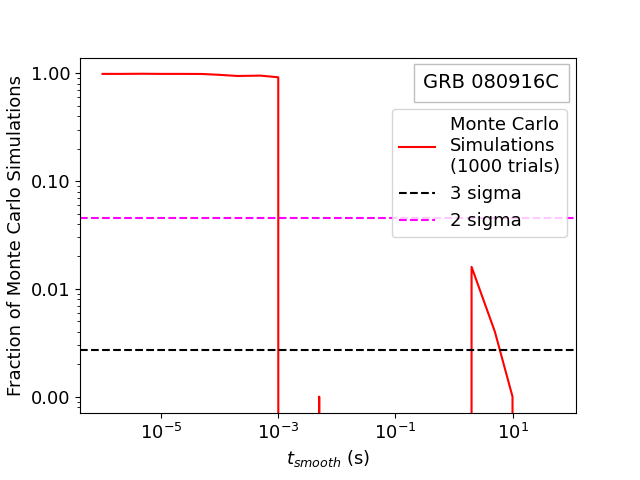}
        \caption[]%
        {{\small GRB 080916C}}    
        \label{080916Cresults}
    \end{subfigure}
    \hfill
    \begin{subfigure}[b]{0.475\textwidth}  
        \centering 
        \includegraphics[width=\textwidth]{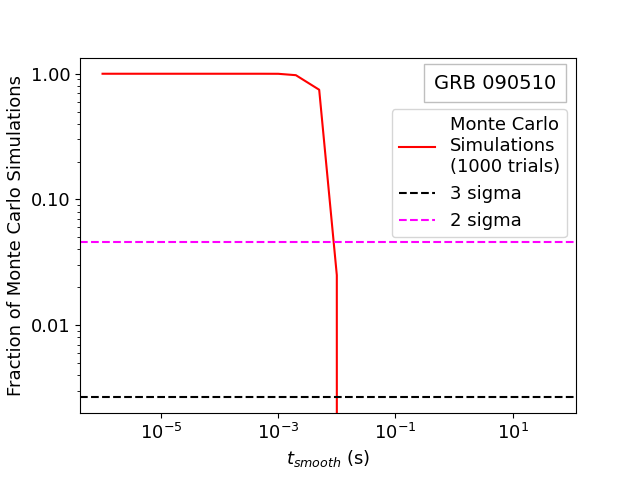}
        \caption[]%
        {{\small GRB 090510}}    
        \label{090510results}
    \end{subfigure}
    
    \begin{subfigure}[b]{0.475\textwidth}   
        \centering 
        \includegraphics[width=\textwidth]{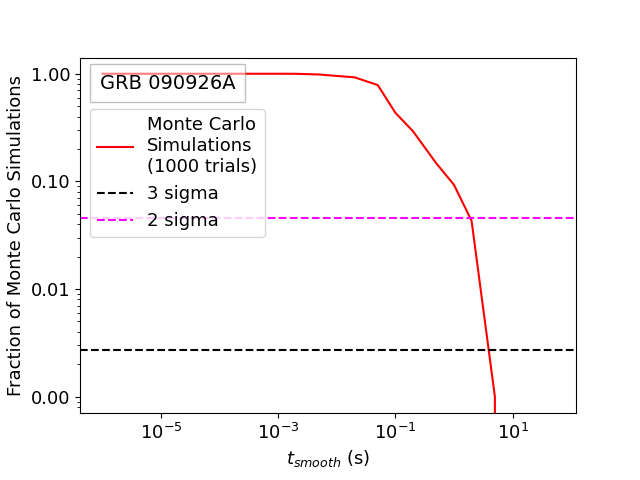}
        \caption[]%
        {{\small GRB 090926A}}    
        \label{131108Aresults}
    \end{subfigure}
    \hfill
    \begin{subfigure}[b]{0.475\textwidth}   
        \centering 
        \includegraphics[width=\textwidth]{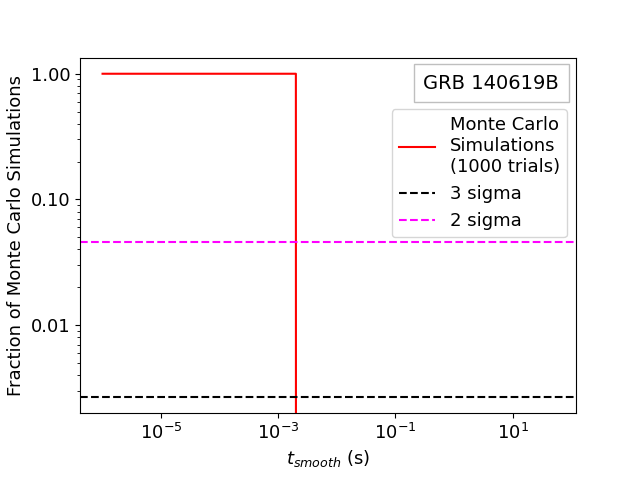}
        \caption[]%
        {{\small GRB 140619B}}    
        \label{140619Bresults}
    \end{subfigure}
    
    \begin{subfigure}[b]{0.475\textwidth}  
        \centering 
        \includegraphics[width=\textwidth]{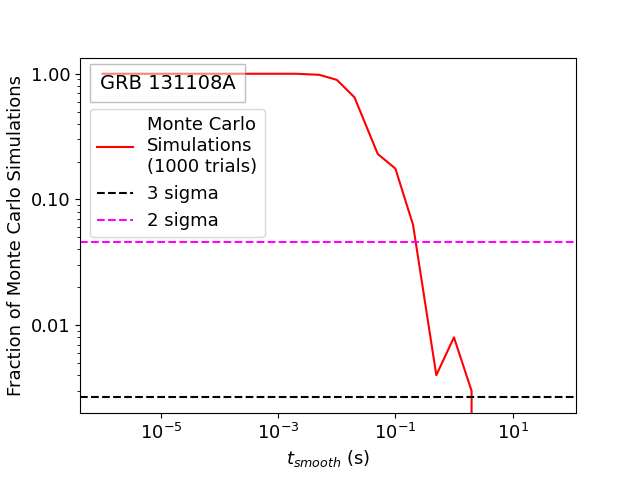}
        \caption[]%
        {{\small GRB 131108A}}    
        \label{131108Aresults2}
    \end{subfigure}
    \hfill
    \begin{subfigure}[b]{0.475\textwidth}   
        \centering 
        \includegraphics[width=\textwidth]{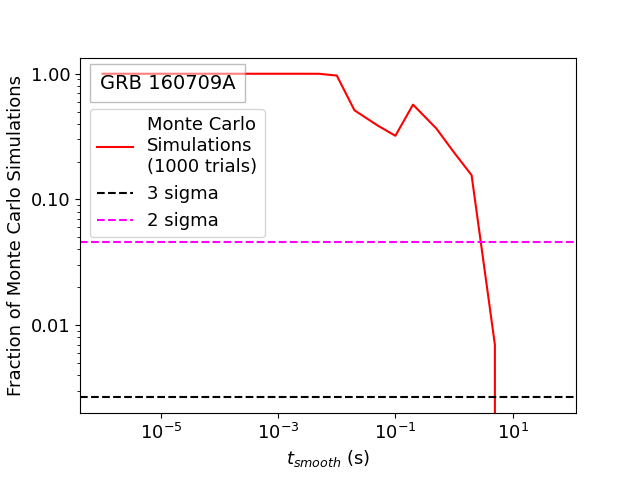}
        \caption[]%
        {{\small GRB 160709A}}    
        \label{160709Aresults}
    \end{subfigure}
    
    \caption[]
    {\small Plots showing the fraction of the 1000 Monte-Carlo GRBs that were bunched on a smaller timescale than the real GRB in the photon pair test. The timescale of smoothing $t_{smooth}$ is shown on the $x$-axis. The fraction plotted is the lowest fraction at any $\Delta t$ value compared.}
    \label{GRBresults}
\end{figure*}

\begin{figure*}
    \centering
    \begin{subfigure}[b]{0.475\textwidth}
        \centering
        \includegraphics[width=\textwidth]{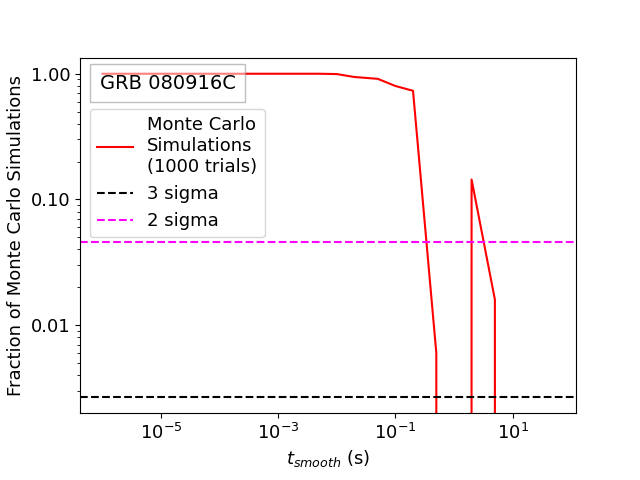}
        \caption[]%
        {{\small GRB 080916C}}    
        \label{080916Cmultdtsresults}
    \end{subfigure}
    \hfill
    \begin{subfigure}[b]{0.475\textwidth}  
        \centering 
        \includegraphics[width=\textwidth]{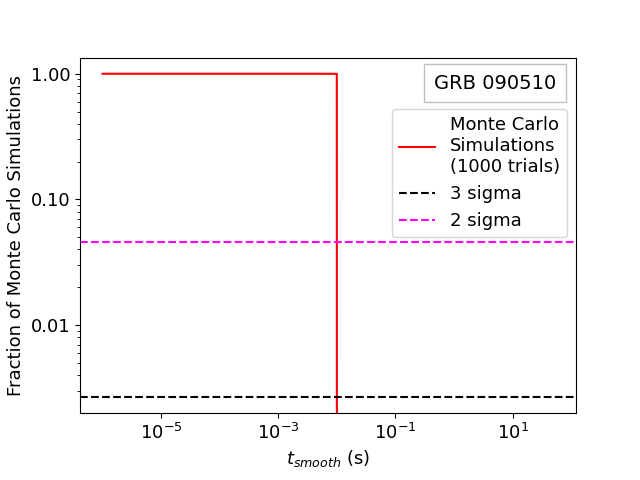}
        \caption[]%
        {{\small GRB 090510}}    
        \label{090510multdtsresults}
    \end{subfigure}
    
    \begin{subfigure}[b]{0.475\textwidth}   
        \centering 
        \includegraphics[width=\textwidth]{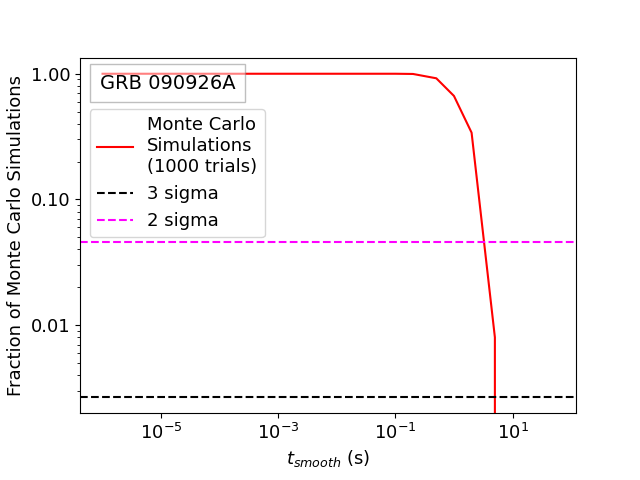}
        \caption[]%
        {{\small GRB 090926A}}    
        \label{090926Amultdtsresults}
    \end{subfigure}
    \hfill
    \begin{subfigure}[b]{0.475\textwidth}   
        \centering 
        \includegraphics[width=\textwidth]{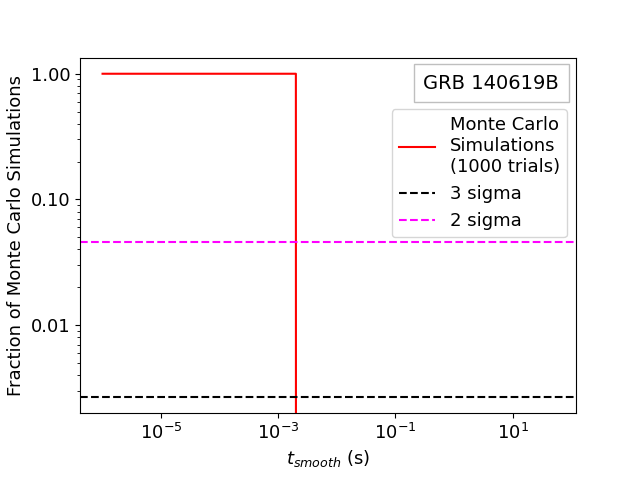}
        \caption[]%
        {{\small GRB 140619B}}    
        \label{140619Bmultdtsresults}
    \end{subfigure}
    
    \begin{subfigure}[b]{0.475\textwidth}  
        \centering 
        \includegraphics[width=\textwidth]{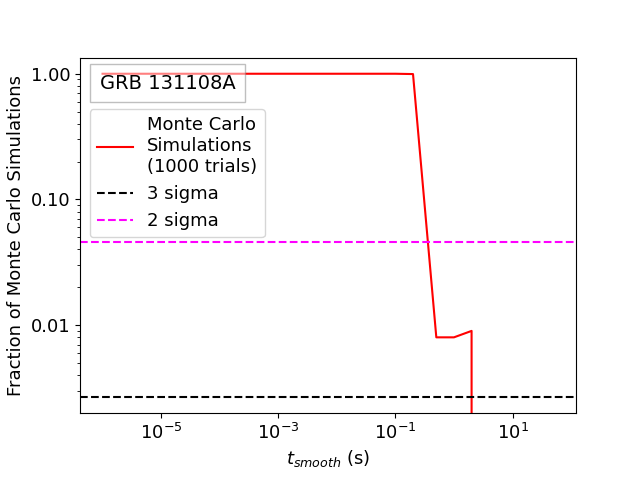}
        \caption[]%
        {{\small GRB 131108A}}    
        \label{131108AAmultdtsresults}
    \end{subfigure}
    \hfill
    \begin{subfigure}[b]{0.475\textwidth}   
        \centering 
        \includegraphics[width=\textwidth]{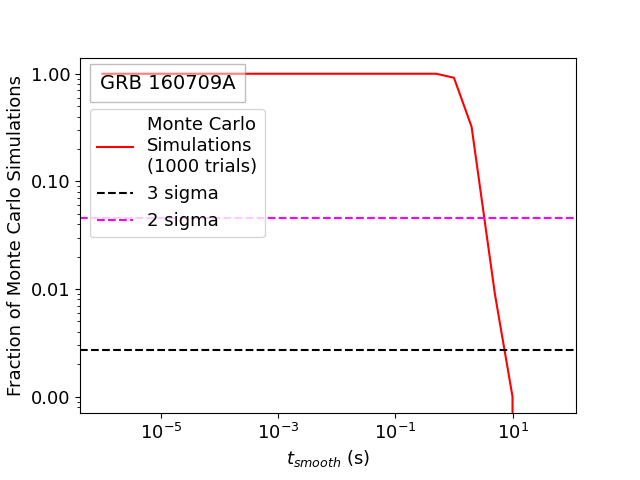}
        \caption[]%
        {{\small GRB 160709A}}    
        \label{160709Amultdtsresults}
    \end{subfigure}
    
    \caption[]
    {\small Plots showing the fraction of the 1000 Monte-Carlo GRBs that were bunched on a smaller timescale than the real GRB according to the time-gap multiplication test. The timescale of smoothing $t_{smooth}$ is shown the $x -$ axis.
    } 
    \label{GRBresults2}
\end{figure*}

\section{Discussion, Summary, \& Conclusions}

The Fermi LAT is particularly useful in the search for short timescales in high-energy GRBs for at least three reasons: first because it measures so many high-energy photons from a GRB as compared to previous missions, second because the LAT records arrival times for individual photons down to the microsecond level, and third because the background at these high energies is so low that most events that occur time and space coincident with a known GRB are likely really photons associated with that GRB -- and not an unrelated background. Additionally, Fermi is useful for detecting multiple GRBs containing over 100 LAT-detected photons, a consequence of having been active now for over 15 years. 

For short GRBs, this study found GRB high-energy variability on timescales at or below 0.01 seconds for two of the short GRBs tested. This confirms the sub-0.01 second variability timescale for GRB 090510 found exclusively at high energies by \citet{2012PhRvL.108w1103N}. A detailed analysis of Fermi data from GRB 140619B was given by \citet{2015ApJ...808..190R}, but no timescale this short was reported at high energies. GRB 160709A is less analyzed (but see \citet{2018JPhCo...2g5013A}) and with no sub-0.01 second variability reported.

In contrast however, for long GRBs, this study found sub-0.01 second high-energy variability for only one of the three long GRBs analyzed -- and then only in one test: specifically the cumulative pair analysis for GRB 080916C. In contrast to \citet{2017A&A...606A..93Y}, our analysis does not immediately confirm the 10-ms variability timescale reported for GRB 090926A at high energies. However, \citet{2017A&A...606A..93Y} reported this variability timescale for only a small portion of this GRB that was particularly fluent at high energies, so the two results could be consistent. Additionally, \citep{2014arXiv1407.0238G} reported a timescale for GRB 131108A below 1-second for the initial peak. Again, our analysis for this GRB was over the entire GRB, and so may be consistent with this GRB sectional result. Stated differently, if sub-0.01 second variability existed only during a small time of the 1000 seconds, this variability might have been ``averaged out" in our analysis by the rest of the data in the complete 1000 seconds studied. 

These results indicate that at least some short GRBs at high energies have a significantly shorter timescale of variability than indicated by the duration of their pulses, in particular their initial pulse. More study is needed for long GRBs.

It is interesting to note that this study was capable of detecting high-energy GRB variability down to the timescale of microseconds -- but did not. That ``only" a super-millisecond timescale was found might be taken as an indication that even shorter timescales don't exist. However, the lack of such a short timescale signal may also indicate that such a signal is so rare that an insufficient number of photons were processed to detect it. It also may indicate that such timescales only existed during a small part of the GRB and was not found with the present analysis because its statistical significance was diminished over the longer parts of the GRB. Future testing may probe smaller timescales by not only analyzing a greater sample of GRBs, but also by analyzing GRBs with greater numbers of high-energy photons, or concentrating on smaller GRB sections in a statistically unbiased manner.

The non-detection of photon counts, particularly at the highest energies where the LAT sensitivity drops, potentially limits the possibility of finding even shorter duration variability at even higher energies, a region of parameter space that is particularly interesting because, in general, GRBs are increasingly variable at higher energies over most of Fermi's energy range \citep{2000ApJ...544..805N}. 

The detected short-timescale signal for short GRBs may have implications regarding the internal physics of GRB photon emission as well as temporally-dispersive properties of the universe between Fermi and the GRB source. As with previously discovered millisecond variability in GRBs, that photons cross the universe in such tight bunches without greater temporal dispersion can be used to probe and limit potential dispersion mechanisms in the intervening universe, including Lorentz-invariance violations, gravitational weak equivalence principle violations, and the nature of pervasive cosmological dark energy.

\section*{Acknowledgements}

We thank Michigan Technological University for support during this research. We thank Oindabi Mukherjee for helpful comments. We acknowledge using ChatGPT to help clarify the language in parts of the text, which was later checked for scientific accuracy by the authors. ChatGPT was also used to take existing code and optimize it for speed, which was also checked for accuracy against pre-optimized code results.

\section*{Data Availability}

The data underlying this article are available generally from the NASA Fermi website at https://fermi.gsfc.nasa.gov/ssc/data/access/ and specifically formatted for the codes used in this article from GitHub at https://github.com/ECaseyAldrich/Rapid-GRB-Variability/tree/main/Photon\%20Lists\%20Used.

\section*{Code Availability}
The codes underlying this article are available from GitHub at https://github.com/ECaseyAldrich/Rapid-GRB-Variability.



\bibliographystyle{mnras}
\bibliography{GrbVariability_MNRAS} 

\begin{thebibliography}{}
\makeatletter
\relax
\def\mn@urlcharsother{\let\do\@makeother \do\$\do\&\do\#\do\^\do\_\do\%\do\~}
\def\mn@doi{\begingroup\mn@urlcharsother \@ifnextchar [ {\mn@doi@}
  {\mn@doi@[]}}
\def\mn@doi@[#1]#2{\def\@tempa{#1}\ifx\@tempa\@empty \href
  {http://dx.doi.org/#2} {doi:#2}\else \href {http://dx.doi.org/#2} {#1}\fi
  \endgroup}
\def\mn@eprint#1#2{\mn@eprint@#1:#2::\@nil}
\def\mn@eprint@arXiv#1{\href {http://arxiv.org/abs/#1} {{\tt arXiv:#1}}}
\def\mn@eprint@dblp#1{\href {http://dblp.uni-trier.de/rec/bibtex/#1.xml}
  {dblp:#1}}
\def\mn@eprint@#1:#2:#3:#4\@nil{\def\@tempa {#1}\def\@tempb {#2}\def\@tempc
  {#3}\ifx \@tempc \@empty \let \@tempc \@tempb \let \@tempb \@tempa \fi \ifx
  \@tempb \@empty \def\@tempb {arXiv}\fi \@ifundefined
  {mn@eprint@\@tempb}{\@tempb:\@tempc}{\expandafter \expandafter \csname
  mn@eprint@\@tempb\endcsname \expandafter{\@tempc}}}

\bibitem[\protect\citeauthoryear{{Ackermann} et~al.,}{{Ackermann}
  et~al.}{2013}]{2013ApJS..209...11A}
{Ackermann} M.,  et~al., 2013, \mn@doi [\apjs] {10.1088/0067-0049/209/1/11},
  \href {https://ui.adsabs.harvard.edu/abs/2013ApJS..209...11A} {209, 11}

\bibitem[\protect\citeauthoryear{{Ackermann} et~al.,}{{Ackermann}
  et~al.}{2014}]{2014Sci...343...42A}
{Ackermann} M.,  et~al., 2014, \mn@doi [Science] {10.1126/science.1242353},
  \href {https://ui.adsabs.harvard.edu/abs/2014Sci...343...42A} {343, 42}

\bibitem[\protect\citeauthoryear{{Ajello} et~al.,}{{Ajello}
  et~al.}{2019}]{2019ApJ...878...52A}
{Ajello} M.,  et~al., 2019, \mn@doi [\apj] {10.3847/1538-4357/ab1d4e}, \href
  {https://ui.adsabs.harvard.edu/abs/2019ApJ...878...52A} {878, 52}

\bibitem[\protect\citeauthoryear{{Aldrich}, {Mukherjee}  \&
  {Nemiroff}}{{Aldrich} et~al.}{2022}]{2022RNAAS...6..237A}
{Aldrich} E.~C.,  {Mukherjee} O.,   {Nemiroff} R.~J.,  2022, \mn@doi [Research
  Notes of the American Astronomical Society] {10.3847/2515-5172/aca15c}, \href
  {https://ui.adsabs.harvard.edu/abs/2022RNAAS...6..237A} {6, 237}

\bibitem[\protect\citeauthoryear{{Atwood} et~al.,}{{Atwood}
  et~al.}{2009}]{2009ApJ...697.1071A}
{Atwood} W.~B.,  et~al., 2009, \mn@doi [\apj] {10.1088/0004-637X/697/2/1071},
  \href {https://ui.adsabs.harvard.edu/abs/2009ApJ...697.1071A} {697, 1071}

\bibitem[\protect\citeauthoryear{{Atwood} et~al.,}{{Atwood}
  et~al.}{2013}]{2013arXiv1303.3514A}
{Atwood} W.,  et~al., 2013, arXiv e-prints, \href
  {https://ui.adsabs.harvard.edu/abs/2013arXiv1303.3514A} {p. arXiv:1303.3514}

\bibitem[\protect\citeauthoryear{{Augusto}, {Navia}, {de Oliveira},
  {Nepomuceno}, {Kopenkin}  \& {Sinzi}}{{Augusto}
  et~al.}{2018}]{2018JPhCo...2g5013A}
{Augusto} C.~R.~A.,  {Navia} C.~E.,  {de Oliveira} M.~N.,  {Nepomuceno} A.,
  {Kopenkin} V.,   {Sinzi} T.,  2018, \mn@doi [Journal of Physics
  Communications] {10.1088/2399-6528/aad3a0}, \href
  {https://ui.adsabs.harvard.edu/abs/2018JPhCo...2g5013A} {2, 075013}

\bibitem[\protect\citeauthoryear{{Bhat}, {Fishman}, {Meegan}, {Wilson}, {Brock}
   \& {Paciesas}}{{Bhat} et~al.}{1992}]{1992Natur.359..217B}
{Bhat} P.~N.,  {Fishman} G.~J.,  {Meegan} C.~A.,  {Wilson} R.~B.,  {Brock}
  M.~N.,   {Paciesas} W.~S.,  1992, \mn@doi [\nat] {10.1038/359217a0}, \href
  {https://ui.adsabs.harvard.edu/abs/1992Natur.359..217B} {359, 217}

\bibitem[\protect\citeauthoryear{{Cucchiara} et~al.,}{{Cucchiara}
  et~al.}{2011}]{2011ApJ...736....7C}
{Cucchiara} A.,  et~al., 2011, \mn@doi [\apj] {10.1088/0004-637X/736/1/7},
  \href {https://ui.adsabs.harvard.edu/abs/2011ApJ...736....7C} {736, 7}

\bibitem[\protect\citeauthoryear{{Daigne}, {Bo{\v{s}}njak}  \&
  {Dubus}}{{Daigne} et~al.}{2011}]{2011A&A...526A.110D}
{Daigne} F.,  {Bo{\v{s}}njak} {\v{Z}}.,   {Dubus} G.,  2011, \mn@doi [\aap]
  {10.1051/0004-6361/201015457}, \href
  {https://ui.adsabs.harvard.edu/abs/2011A&A...526A.110D} {526, A110}

\bibitem[\protect\citeauthoryear{{Desai}}{{Desai}}{1981}]{1981Ap&SS..75...15D}
{Desai} U.~D.,  1981, \mn@doi [\apss] {10.1007/BF00651380}, \href
  {https://ui.adsabs.harvard.edu/abs/1981Ap&SS..75...15D} {75, 15}

\bibitem[\protect\citeauthoryear{{Eichler}, {Livio}, {Piran}  \&
  {Schramm}}{{Eichler} et~al.}{1989}]{1989Natur.340..126E}
{Eichler} D.,  {Livio} M.,  {Piran} T.,   {Schramm} D.~N.,  1989, \mn@doi
  [\nat] {10.1038/340126a0}, \href
  {https://ui.adsabs.harvard.edu/abs/1989Natur.340..126E} {340, 126}

\bibitem[\protect\citeauthoryear{Epanechnikov}{Epanechnikov}{1969}]{KDFarticle}
Epanechnikov V.~A.,  1969, \mn@doi [Theory of Probability \& Its Applications]
  {10.1137/1114019}, 14, 153

\bibitem[\protect\citeauthoryear{Fermi-Collaboration}{Fermi-Collaboration}{2023}]{WebSiteFermi}
Fermi-Collaboration 2023, Overview of the LAT,
  \url{https://fermi.gsfc.nasa.gov/ssc/data/analysis/documentation/Cicerone/Cicerone_Introduction/LAT_overview.html}

\bibitem[\protect\citeauthoryear{{Gao}, {Wu}  \& {M{\'e}sz{\'a}ros}}{{Gao}
  et~al.}{2015}]{2015ApJ...810..121G}
{Gao} H.,  {Wu} X.-F.,   {M{\'e}sz{\'a}ros} P.,  2015, \mn@doi [\apj]
  {10.1088/0004-637X/810/2/121}, \href
  {https://ui.adsabs.harvard.edu/abs/2015ApJ...810..121G} {810, 121}

\bibitem[\protect\citeauthoryear{{Giuliani} et~al.,}{{Giuliani}
  et~al.}{2014}]{2014arXiv1407.0238G}
{Giuliani} A.,  et~al., 2014, \mn@doi [arXiv e-prints]
  {10.48550/arXiv.1407.0238}, \href
  {https://ui.adsabs.harvard.edu/abs/2014arXiv1407.0238G} {p. arXiv:1407.0238}

\bibitem[\protect\citeauthoryear{Greiner}{Greiner}{2021}]{GreinerWebCat}
Greiner J.~C.,  2021, GRBs localized, \url
  {https://www.mpe.mpg.de/~jcg/grbgen.html}

\bibitem[\protect\citeauthoryear{{Hasco{\"e}t}, {Daigne}, {Mochkovitch}  \&
  {Vennin}}{{Hasco{\"e}t} et~al.}{2012}]{2012MNRAS.421..525H}
{Hasco{\"e}t} R.,  {Daigne} F.,  {Mochkovitch} R.,   {Vennin} V.,  2012,
  \mn@doi [\mnras] {10.1111/j.1365-2966.2011.20332.x}, \href
  {https://ui.adsabs.harvard.edu/abs/2012MNRAS.421..525H} {421, 525}

\bibitem[\protect\citeauthoryear{Härdle, Müller, Sperlich  \&
  Werwatz}{Härdle et~al.}{2006}]{KDEbook}
Härdle W.~K.,  Müller M.,  Sperlich S.,   Werwatz A.,  2006, Nonparametric
  and Semiparametric Models, \mn@doi{10.1007/978-3-642-17146-8.
}

\bibitem[\protect\citeauthoryear{{Kouveliotou}, {Meegan}, {Fishman}, {Bhat},
  {Briggs}, {Koshut}, {Paciesas}  \& {Pendleton}}{{Kouveliotou}
  et~al.}{1993}]{1993ApJ...413L.101K}
{Kouveliotou} C.,  {Meegan} C.~A.,  {Fishman} G.~J.,  {Bhat} N.~P.,  {Briggs}
  M.~S.,  {Koshut} T.~M.,  {Paciesas} W.~S.,   {Pendleton} G.~N.,  1993,
  \mn@doi [\apjl] {10.1086/186969}, \href
  {https://ui.adsabs.harvard.edu/abs/1993ApJ...413L.101K} {413, L101}

\bibitem[\protect\citeauthoryear{{Levan}}{{Levan}}{2018}]{2018grb..book.....L}
{Levan} A.,  2018, {Gamma-Ray Bursts}, \mn@doi{10.1088/2514-3433/aae164.
}

\bibitem[\protect\citeauthoryear{{MacLachlan}, {Shenoy}, {Sonbas}, {Dhuga},
  {Eskandarian}, {Maximon}  \& {Parke}}{{MacLachlan}
  et~al.}{2012}]{2012MNRAS.425L..32M}
{MacLachlan} G.~A.,  {Shenoy} A.,  {Sonbas} E.,  {Dhuga} K.~S.,  {Eskandarian}
  A.,  {Maximon} L.~C.,   {Parke} W.~C.,  2012, \mn@doi [\mnras]
  {10.1111/j.1745-3933.2012.01295.x}, \href
  {https://ui.adsabs.harvard.edu/abs/2012MNRAS.425L..32M} {425, L32}

\bibitem[\protect\citeauthoryear{{MacLachlan} et~al.,}{{MacLachlan}
  et~al.}{2013}]{2013MNRAS.432..857M}
{MacLachlan} G.~A.,  et~al., 2013, \mn@doi [\mnras] {10.1093/mnras/stt241},
  \href {https://ui.adsabs.harvard.edu/abs/2013MNRAS.432..857M} {432, 857}

\bibitem[\protect\citeauthoryear{{Nemiroff}}{{Nemiroff}}{2000}]{2000ApJ...544..805N}
{Nemiroff} R.~J.,  2000, \mn@doi [\apj] {10.1086/317230}, \href
  {https://ui.adsabs.harvard.edu/abs/2000ApJ...544..805N} {544, 805}

\bibitem[\protect\citeauthoryear{{Nemiroff}}{{Nemiroff}}{2012}]{2012MNRAS.419.1650N}
{Nemiroff} R.~J.,  2012, \mn@doi [\mnras] {10.1111/j.1365-2966.2011.19838.x},
  \href {https://ui.adsabs.harvard.edu/abs/2012MNRAS.419.1650N} {419, 1650}

\bibitem[\protect\citeauthoryear{{Nemiroff}, {Connolly}, {Holmes}  \&
  {Kostinski}}{{Nemiroff} et~al.}{2012}]{2012PhRvL.108w1103N}
{Nemiroff} R.~J.,  {Connolly} R.,  {Holmes} J.,   {Kostinski} A.~B.,  2012,
  \mn@doi [\prl] {10.1103/PhysRevLett.108.231103}, \href
  {https://ui.adsabs.harvard.edu/abs/2012PhRvL.108w1103N} {108, 231103}

\bibitem[\protect\citeauthoryear{{Norris}, {Bonnell}, {Nemiroff}, {Scargle},
  {Kouveliotou}, {Paciesas}, {Meegan}  \& {Fishman}}{{Norris}
  et~al.}{1995}]{1995ApJ...439..542N}
{Norris} J.~P.,  {Bonnell} J.~T.,  {Nemiroff} R.~J.,  {Scargle} J.~D.,
  {Kouveliotou} C.,  {Paciesas} W.~S.,  {Meegan} C.~A.,   {Fishman} G.~J.,
  1995, \mn@doi [\apj] {10.1086/175194}, \href
  {https://ui.adsabs.harvard.edu/abs/1995ApJ...439..542N} {439, 542}

\bibitem[\protect\citeauthoryear{{Norris}, {Nemiroff}, {Bonnell}, {Scargle},
  {Kouveliotou}, {Paciesas}, {Meegan}  \& {Fishman}}{{Norris}
  et~al.}{1996}]{1996ApJ...459..393N}
{Norris} J.~P.,  {Nemiroff} R.~J.,  {Bonnell} J.~T.,  {Scargle} J.~D.,
  {Kouveliotou} C.,  {Paciesas} W.~S.,  {Meegan} C.~A.,   {Fishman} G.~J.,
  1996, \mn@doi [\apj] {10.1086/176902}, \href
  {https://ui.adsabs.harvard.edu/abs/1996ApJ...459..393N} {459, 393}

\bibitem[\protect\citeauthoryear{{Nusser}}{{Nusser}}{2016}]{2016ApJ...821L...2N}
{Nusser} A.,  2016, \mn@doi [\apjl] {10.3847/2041-8205/821/1/L2}, \href
  {https://ui.adsabs.harvard.edu/abs/2016ApJ...821L...2N} {821, L2}

\bibitem[\protect\citeauthoryear{{Piran}}{{Piran}}{2004}]{2004RvMP...76.1143P}
{Piran} T.,  2004, \mn@doi [Reviews of Modern Physics]
  {10.1103/RevModPhys.76.1143}, \href
  {https://ui.adsabs.harvard.edu/abs/2004RvMP...76.1143P} {76, 1143}

\bibitem[\protect\citeauthoryear{{Preece}, {Briggs}, {Mallozzi}, {Pendleton},
  {Paciesas}  \& {Band}}{{Preece} et~al.}{1998}]{1998ApJ...506L..23P}
{Preece} R.~D.,  {Briggs} M.~S.,  {Mallozzi} R.~S.,  {Pendleton} G.~N.,
  {Paciesas} W.~S.,   {Band} D.~L.,  1998, \mn@doi [\apjl] {10.1086/311644},
  \href {https://ui.adsabs.harvard.edu/abs/1998ApJ...506L..23P} {506, L23}

\bibitem[\protect\citeauthoryear{{Ruffini} et~al.,}{{Ruffini}
  et~al.}{2015}]{2015ApJ...808..190R}
{Ruffini} R.,  et~al., 2015, \mn@doi [\apj] {10.1088/0004-637X/808/2/190},
  \href {https://ui.adsabs.harvard.edu/abs/2015ApJ...808..190R} {808, 190}

\bibitem[\protect\citeauthoryear{{Ryde} \& {Svensson}}{{Ryde} \&
  {Svensson}}{2002}]{2002ApJ...566..210R}
{Ryde} F.,  {Svensson} R.,  2002, \mn@doi [\apj] {10.1086/337962}, \href
  {https://ui.adsabs.harvard.edu/abs/2002ApJ...566..210R} {566, 210}

\bibitem[\protect\citeauthoryear{{Sarkar}}{{Sarkar}}{2002}]{2002MPLA...17.1025S}
{Sarkar} S.,  2002, \mn@doi [Modern Physics Letters A]
  {10.1142/S0217732302007521}, \href
  {https://ui.adsabs.harvard.edu/abs/2002MPLA...17.1025S} {17, 1025}

\bibitem[\protect\citeauthoryear{{Schaefer}}{{Schaefer}}{1999}]{1999PhRvL..82.4964S}
{Schaefer} B.~E.,  1999, \mn@doi [\prl] {10.1103/PhysRevLett.82.4964}, \href
  {https://ui.adsabs.harvard.edu/abs/1999PhRvL..82.4964S} {82, 4964}

\bibitem[\protect\citeauthoryear{{Shahmoradi} \& {Nemiroff}}{{Shahmoradi} \&
  {Nemiroff}}{2015}]{2015MNRAS.451..126S}
{Shahmoradi} A.,  {Nemiroff} R.~J.,  2015, \mn@doi [\mnras]
  {10.1093/mnras/stv714}, \href
  {https://ui.adsabs.harvard.edu/abs/2015MNRAS.451..126S} {451, 126}

\bibitem[\protect\citeauthoryear{{Tangmatitham} \& {Nemiroff}}{{Tangmatitham}
  \& {Nemiroff}}{2019}]{2019arXiv190305688T}
{Tangmatitham} M.,  {Nemiroff} R.~J.,  2019, arXiv e-prints, \href
  {https://ui.adsabs.harvard.edu/abs/2019arXiv190305688T} {p. arXiv:1903.05688}

\bibitem[\protect\citeauthoryear{Triveri}{Triveri}{2023}]{CodeCiteTriveri}
Triveri J.,  2023, Kernel Density Estimation from Scratch in Python,
  \url{https://www.jtrive.com/kernel-density-estimation-from-scratch-in-python.html}

\bibitem[\protect\citeauthoryear{{Walker}, {Schaefer}  \& {Fenimore}}{{Walker}
  et~al.}{2000}]{2000ApJ...537..264W}
{Walker} K.~C.,  {Schaefer} B.~E.,   {Fenimore} E.~E.,  2000, \mn@doi [\apj]
  {10.1086/308995}, \href
  {https://ui.adsabs.harvard.edu/abs/2000ApJ...537..264W} {537, 264}

\bibitem[\protect\citeauthoryear{{Woosley} \& {Bloom}}{{Woosley} \&
  {Bloom}}{2006}]{2006ARA&A..44..507W}
{Woosley} S.~E.,  {Bloom} J.~S.,  2006, \mn@doi [\araa]
  {10.1146/annurev.astro.43.072103.150558}, \href
  {https://ui.adsabs.harvard.edu/abs/2006ARA&A..44..507W} {44, 507}

\bibitem[\protect\citeauthoryear{{Yassine}, {Piron}, {Mochkovitch}  \&
  {Daigne}}{{Yassine} et~al.}{2017}]{2017A&A...606A..93Y}
{Yassine} M.,  {Piron} F.,  {Mochkovitch} R.,   {Daigne} F.,  2017, \mn@doi
  [\aap] {10.1051/0004-6361/201630353}, \href
  {https://ui.adsabs.harvard.edu/abs/2017A&A...606A..93Y} {606, A93}

\bibitem[\protect\citeauthoryear{{Zitouni}, {Guessoum}, {AlQassimi}  \&
  {Alaryani}}{{Zitouni} et~al.}{2018}]{2018Ap&SS.363..223Z}
{Zitouni} H.,  {Guessoum} N.,  {AlQassimi} K.~M.,   {Alaryani} O.,  2018,
  \mn@doi [\apss] {10.1007/s10509-018-3449-0}, \href
  {https://ui.adsabs.harvard.edu/abs/2018Ap&SS.363..223Z} {363, 223}

\makeatother
\end{thebibliography}






\bsp	
\label{lastpage}
\end{document}